# Spatially modulated morphotropic phase boundaries in a compressively strained multiferroic thin film

*Ting-Ran Liu[1,†], Xiangwei Guo[2,†], Sajid Husain[3], Maya Ramesh[4], Pushpendra Gupta[3], Darrell Schlom[4], Ramamoorthy Ramesh[3], and Yu-Tsun Shao[1,5]*

[1]Mork Family Department of Chemical Engineering and Materials Science, University of Southern California, Los Angeles, CA 90089, USA.
[2]Department of Mechanical Engineering, University of Michigan, Dearborn, MI 48128, USA.
[3]Department of Materials Science and Engineering, University of California, Berkeley, CA 94720, USA.
[4]Department of Materials Science and Engineering, Cornell University, Ithaca, NY 14850, USA.
[5]Core Center of Excellence in Nano Imaging, University of Southern California, Los Angeles, CA 90089, USA.

[†] T.R.L. and X.G. contributed equally to this work.
*To whom correspondence may be addressed: yutsunsh@usc.edu

## Abstract

The coexisting rhombohedral-like (R', $M_A$) and tetragonal-like (T', $M_C$) monoclinic phases in compressively strained bismuth ferrite thin films exhibit exceptional piezoelectric and magnetic properties. While previous studies have largely focused on probing the morphotropic phase boundaries (MPBs) comprising ordered R'/T' twins, their self-organizing structures remain not fully explored. Here, we observed two types of interphase boundaries in a 60 nm-thick $BiFeO_3$ film epitaxially grown on a $LaAlO_3$ substrate by employing multi-modal diffraction-based electron microscopy. First, the flat MPBs form lines extending >1 mm, and repeat almost every ~20 µm. Additionally, we uncover a new type of phase boundary with zig-zag regions of alternating R'/R' and T'/T' twin domains. Cross-sectional multislice electron ptychography confirms the atomic-scale polarization rotation across the MPB, with out-of-plane strain varying >15%. Plan-view electron backscatter diffraction reveals the lattice disclination of ~1.5° across the zig-zag interphase boundaries, while having >2.5° within the MPB. Phase-field modeling suggests that the formation of zig-zag phase boundaries arises from balancing between Landau and elastic energies. We speculate that such well-ordered interphase boundaries are associated with mesoscale in-plane strain modulations, thus providing a way to engineer and harness their properties for potential device applications.

**Keywords:** multiferroic thin film, morphotropic phase boundary, strain engineering, multislice electron ptychography, scanning transmission electron microscopy, phase-field simulations, electron backscatter diffraction

## 1. Introduction

Phase separation, coexistence, and competition between nearly degenerate states in functional materials often give rise to exceptional physical properties. One notable example is the morphotropic phase boundary (MPB) in lead-based relaxor ferroelectrics such as $Pb(Zr_x,Ti_{1-x})O_3$ (PZT), $Pb(Mg_{0.33},Nb_{0.67})O_3$-$PbTiO_3$ (PMN-PT), and $Pb(Zn_{0.33},Nb_{0.67})O_3$-$PbTiO_3$ (PZN-PT) systems[1,2]. At compositions near the rhombohedral-to-tetragonal boundary, the free energy landscape flattens and the crystallographic symmetry lowers, facilitating polarization rotation and resulting in large dielectric and piezoelectric responses. Besides compositional tuning, phase coexistence can also be stabilized and controlled through epitaxial strain in lead-free systems, including $NaNbO_3$ [3] and $BiFeO_3$ [4] thin films.

Among lead-free ferroelectrics, $BiFeO_3$ (BFO) thin films have been extensively studied due to their room-temperature multiferroicity, which exhibits strong coupling among lattice strain, polarization, and magnetic order. Bulk BFO adopts a rhombohedral ground state ($R3c$, space group No. 161) with a large spontaneous polarization of ~100 $\mu C/cm^2$ oriented along the $<111>_{pc}$ direction[5–7]. Under large (>4%) epitaxial compressive strain, BFO undergoes a structural phase transition to monoclinically distorted, rhombohedral-like (R', $M_A$) and tetragonal-like phases (T', $M_C$), with polarization along the $[uuv]_{pc}$ and $[u0v]_{pc}$ directions, respectively (**Figure 1a**)[8]. Both phases display pronounced tetragonal distortion in response to compressive strain, with tetragonality $c/a$ ~1.08 for R' and $c/a$ ~1.25 for T' (where $c$ and $a$ denote out-of-plane and in-plane lattice constants, respectively). Furthermore, at specific film thicknesses and epitaxial conditions, BFO films often exhibit complex nanoscale stripe-like patterns comprising mixed R' and T' phases[4,9]. Within such nanoscale phase mixtures, both polarization orientation and elastic strain vary drastically across R'/T' interfaces, giving rise to enhanced spontaneous magnetism[10,11], electromechanical response[12], shape-memory effects[13], and electrical conduction along mixed-phase boundaries[14].

A key requirement for understanding strain-driven MPBs in BFO thin films is determining how coexisting R' and T' phases, as well as nanoscale phase mixtures, organize across the film at mesoscopic length scales. Prior studies have largely focused on phase fractions controlled by boundary conditions, electric fields, or film thickness, as well as resolving atomic-scale interphase boundary structures[15,16]. While these efforts have provided critical insight into local phase behavior and interfacial physics, they have primarily concentrated on stripe-like mixed-phase regions of the film. As a result, whether R' and T' phases self-organize into ordered mesoscale patterns under epitaxial constraint, and how such organization relates to long-range strain accommodation, remain largely unexplored.

Here, we identify two distinct types of interphase boundaries in a compressively strained BFO thin film: a previously unreported zigzag boundary formed by alternating R'/R' and T'/T' twin domains, and the well-studied stripe-like R'/T' mixed-phase boundary. In this manuscript, we will refer to them as "zigzag" and "MPB" interphase boundaries, respectively. Using multi-modal diffraction-based electron microscopy, we characterize these interphase boundaries across atomic, nanometer, and mesoscopic length scales. Atomic-resolution imaging reveals continuous polarization rotation and large strain gradients across the R'/T' boundary, while local phase fractions indicate strain relaxation within the mixed-phase regions. Mesoscale diffraction mapping further shows that both stripe-like and zigzag interphase boundaries are associated with pronounced lattice disclinations, and extend continuously over length scales exceeding 500 µm. Building on the established strain-driven MPB phase diagram[4,17–19], this work identifies nano- and mesoscale organization of interphase boundaries and establishes a mesoscale pathway for enhanced electromechanical response in BFO films.

## 2. Results

We examined R' and T' interphase boundaries in a 60 nm-thick epitaxially strained $BiFeO_3$ film grown on a (001) $LaAlO_3$ substrate with a 10 nm (La, Sr)$MnO_3$ bottom electrode. X-ray diffraction (XRD) **(Figure S1)** and atomic force microscopy (AFM) results (**Figure 1b**) confirm the coexistence of intermixed phases. To resolve the atomic structure within the phase mixture regions, we performed multi-slice electron ptychography (MEP). MEP reconstruction solves the inverse electron-scattering problem to reconstruct the electrostatic potential, enabling quantitative structural analysis of both cation and anion sublattices with sub-30 pm lateral resolution and precision of ~1 pm[20,21]. From the reconstructed MEP images, quantitative information including polarization rotation, lattice rotation and disclinations can be extracted.

The polarization vectors in the R' and T' phases differ in both direction and magnitude, and rotate continuously across the interphase boundaries (**Figure 1c**). By measuring the displacement between the geometric centers of Bi and Fe atomic columns, the polar displacements are determined to be 31 pm for R' and 51 pm for T' phases, with both polarization vectors oriented within the *xz*-plane, consistent with their respective monoclinic symmetries. The R' phase exhibits a larger in-plane *x*-component of polarization, resulting in its orientation tilted further away from the substrate normal and closer to the $<101>_{pc}$ direction, consistent with $M_A$ symmetry. In contrast, the T' phase shows a reduced in-plane component and a dominant out-of-plane contribution, yielding a polarization direction closer to $<001>_{pc}$, consistent with $M_C$ symmetry.

We further determined the unit-cell tetragonality (*c/a* ratio, where *c* and *a* denote the out-of-plane and in-plane lattice parameters, respectively) by measuring the distances between neighboring Bi atomic columns. As shown in **Figure 1d**, the interphase boundary is clearly resolved by the distinct tetragonalities of the two phases. The average *c/a* ratios are measured to be ~1.25 for T' and ~1.07 for R' phases, consistent with previous measurements with XRD measurements[22] **(Figure S1)**. In addition to polarization direction rotation across the interphase boundaries, the unit cells exhibit a lattice rotation of ~3.8° with respect to the [010]$_{pc}$ direction when transitioning from the R' to the T' region (**Figure 1e**). This lattice rotation results from the elastic incompatibility between the R' and T' unit cells. Moreover, elongated atomic columns and dumbbell-like features **(Figure S2)** are observed near the interphase boundaries in the MEP reconstructions. Because MEP reconstruction provides depth sensitivity, these elongated features correspond to atomic displacements along the projection direction, indicating the presence of an out-of-plane lattice disclination of ~2.2°.

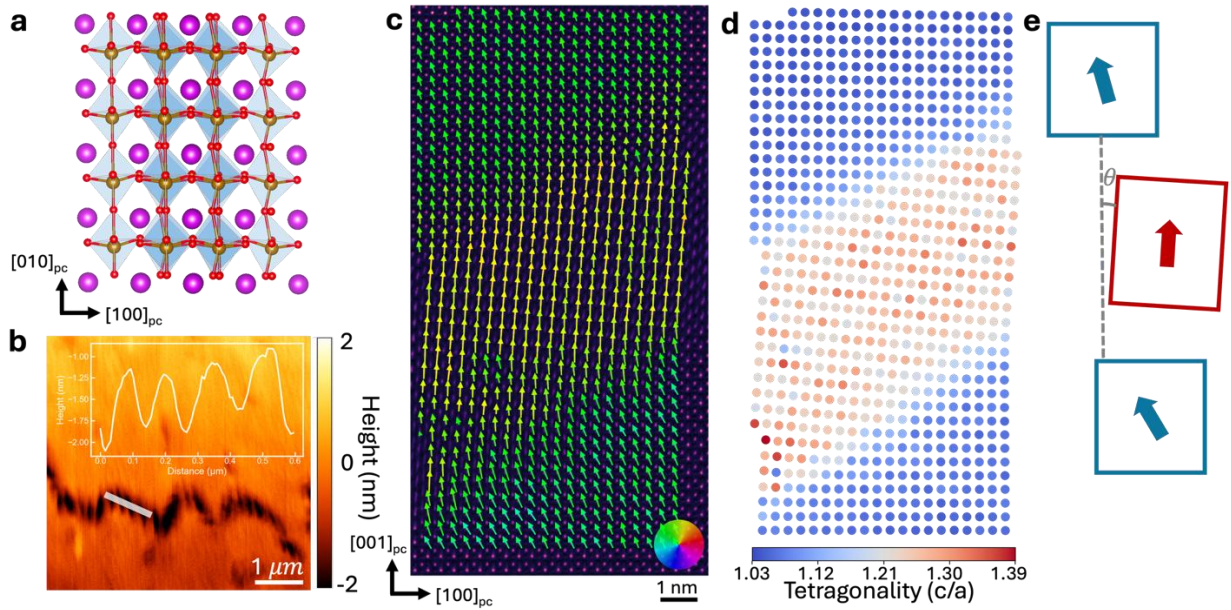

**Figure 1. Atomic structures of the MPB with coexisting rhombohedral-like (R') and tetragonal-like (T') phases in compressively strained BFO thin film. (a)** Structure model of T' phase BFO with monoclinic symmetry and large tetragonal distortion. **(b)** AFM scan of the coexisting R'/T' phase mixture region. The inset shows the height line profile of the white line. **(c)** Cross-sectional MEP reconstruction of the MPB region. Polarization rotation can be observed across the phase boundary. **(d)** Unit cell tetragonality c/a extracted from the MEP reconstruction, showing c/a ~1.08 for R' and 1.25 for T' phases. The c and a denote out-of-plane and in-plane lattice parameters, respectively. **(e)** Schematic showing the unit cell rotation across the boundary, where blue and red represent unit cells in R' and T' BFO. The arrows represent polarization direction in both phases. There is a relative unit cell rotation of ~3.8° between the two phases. Scale bar in (c) represents 1 nm.

To explore the nanoscale-to-mesoscale organization of the R' and T' phases, we employed two electron diffraction-based techniques, depth-resolved electron diffraction imaging (DREDI)[23] and electron backscatter diffraction (EBSD), both implemented in widely accessible scanning electron microscopes (SEMs). These techniques record crystallographic information in reciprocal space using either a segmented detector (DREDI) or a CMOS detector (EBSD), enabling spatially resolved phase mapping over large areas. In both methods, image contrast arises from changes in the intensity and geometry of Kikuchi diffraction patterns, which are formed primarily by thermally diffuse-scattered electrons that subsequently undergo Bragg diffraction[24–27]. Because Kikuchi patterns are governed by dynamical diffraction processes, their intensity distributions are highly sensitive to local crystal symmetry, lattice distortion, and orientation. Consequently, subtle differences between the monoclinic R' ($M_A$) and T' ($M_C$) phases produce distinct Kikuchi intensities, allowing reliable phase discrimination and mapping across nano- to mesoscopic scales.

DREDI reveals two types of interphase boundaries, flat MPBs and zigzag boundaries, which extend continuously over 500 μm. These interphase boundaries enclose ~20 μm-wide superdomains composed of either T'/T' or R'/R' ferroelastic twins. As shown in **Figure 2(a, b)**, the DREDI image exhibits alternating bright and dark contrast, reflecting quasi-periodic changes in the diffraction condition arising from lattice disclinations. The flat, millimeter-long MPBs (marked by white boxes in the DREDI image) recur quasi-periodically with a spacing of ~20 $\mu m$ **(Figure S3)**. In addition to these flat MPBs, we observed a second class of interphase boundaries characterized by frustrated, zigzag patterns, which also appear quasi-periodically across the film.

To further understand the structural details across these two types of interphase boundaries, we employed EBSD to capture the full diffraction-space information. EBSD records a two-dimensional Kikuchi pattern ($k_x$, $k_y$) at each real-space probe position ($x$, $y$), yielding four-dimensional (4D) datasets. We applied k-means clustering to the EBSD data to group real-space regions based on similarities in their corresponding Kikuchi patterns. Within the MPB region, two distinct clusters can be identified, marked in cyan and magenta in **Figure 2c**. The corresponding Kikuchi patterns show two crystallographically distinct phases, with differences in specific Kikuchi bands highlighted by yellow arrows in **Figure 2d**. An overlay of two sets of diffraction patterns from the two clusters show clear band splitting and shifts associated with one of the phases (cyan). Quantitative analysis of the Kikuchi bands shifts enables the construction of lattice misorientation maps from the EBSD datasets (**Figure 2e**), which reveal lattice disclinations of ~1.5° across the zig-zag interphase boundaries, and >2.5° within the MPB regions. The periodic and alternating occurrence of these two interphase boundary types indicates a laterally modulated lattice deformation that extends from mesoscale length scales to the macroscopic dimensions, spanning the whole 3 mm-wide film.

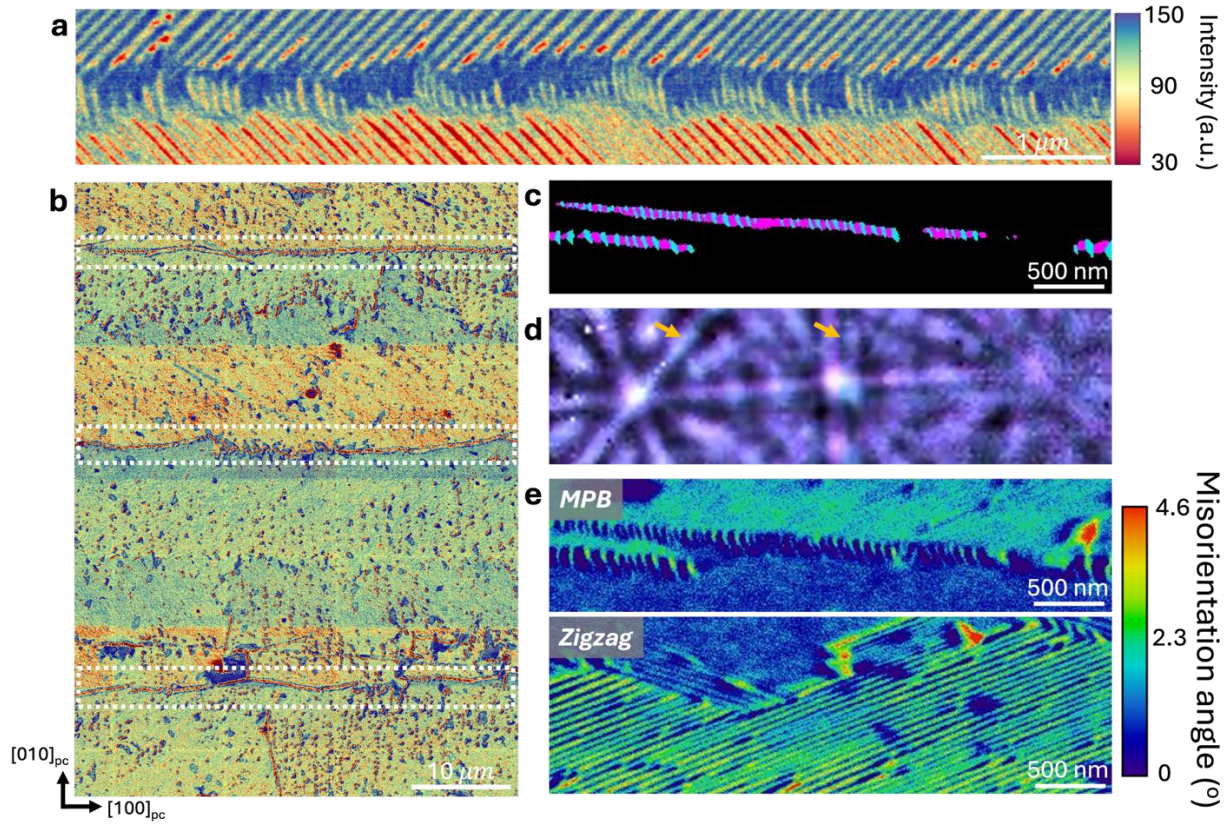

**Figure 2. Distinct types of laterally-ordered R' and T' interphase boundaries. (a)** DREDI image of a flat MPB region. **(b)** DREDI image over 60 μm field-of-view along plane-view (*xy*-plane). The alternating contrast arising from distinct diffraction intensities reflects the crystallographic symmetries and lattice disclination for each phase. The interphase boundaries extend over 500 μm. The flat MPBs (white boxes) appear with ~20 μm intervals. **(c)** EBSD was employed to further explore the full momentum information captured by a CMOS detector. False-colored MPB domain structure constructed from k-means clustering. The cyan and magenta domains represent two distinct backscattered diffraction patterns. **(d)** Overlay of electron backscattered diffraction patterns from two phases within MPB in (b) with the same color coding. The Kikuchi band splitting and shifts between cyan and magenta domains (marked by yellow arrows) indicate the distinct crystal symmetries and lattice misorientation. **(e)** Misorientation maps from representative EBSD data showing two types of interfaces present in the sample: a flat, well-aligned MPB and a frustrated, zigzag phase boundary.

To resolve the strain distribution within the periodically occurring MPB regions, we performed cross-sectional scanning electron nanodiffraction (SEND). SEND records a convergent beam electron diffraction (CBED) pattern at each probe position, yielding a 4D dataset. By quantitatively measuring shifts in the Bragg reflections, SEND enables quantitative strain mapping in the cross-sectional *xz*-plane, providing information complementary to the plan-view lattice disclination maps obtained from EBSD **(Figure 2)**. To further reduce common artifacts arising from lattice mistilt and thickness variations, we applied exit-wave power Cepstrum (EWPC) analysis[28]. Using the LAO substrate as a reference, we mapped the local strain distribution within the BFO thin film.

**Figure 3** shows the SEND analysis of the *xz*-plane, viewed normal to the MPB extension direction. Stripe-like diffraction contrast in virtual annular dark-field (ADF) image indicates the coexistence of R' and T' phases within the MPB region (**Figure 3a**). The CBED pattern extracted from the yellow box (**Figure 3b**) exhibits a clear off-axis lattice disclination of ~2.2° away from the zone axis. **Figures 3c and 3d** show maps of in-plane and out-of-plane strain components (uniaxial tensile $\varepsilon_{xx}$ and $\varepsilon_{zz}$), corresponding to changes in the *a*- and *c*-axis lattice parameters within the MPB. The T' phase exhibits a maximum out-of-plane tensile strain of ~15% (**Figure 3d**), consistent with reported values[29,30]. In addition, the nonuniform strain distribution within the T' phase suggests the presence of additional lattice deformation such as bending. The lattice rotation map in the *xz*-plane further confirms a ~3.8° rotation between the R' and T' phases, consistent with the interphase unit-cell rotation observed from the MEP reconstructions **(Figure 1e)**. Line profiles across the MPB (**Figure 3f**) show clear differences in lattice constants along the *c*-axis (green) and *a*-axis (yellow). Such drastic strain variation is expected in an MPB, as the coexisting R' and T' phases are mechanically incompatible[18,31]. Notably, the strain maps indicate that the MPB region is predominantly R' rich, occupying an area fraction of ~93.6% **(Figure 2c)**, corresponding to an effective compressive strain of ~3.1% in the reported strain-phase diagram[18]. Although the nominal compressive strain of 4% is imposed by the film-substrate lattice mismatch, the R'-rich MPB suggests local strain relaxation. Given that MPBs recur quasi-periodically at ~20 µm intervals, we can expect that other regions may exhibit locally larger compressive strain, forming mesoscopic bands of strain modulation across the entire film. We speculate that such strain modulations may promote the formation of zigzag interphase boundaries[9].

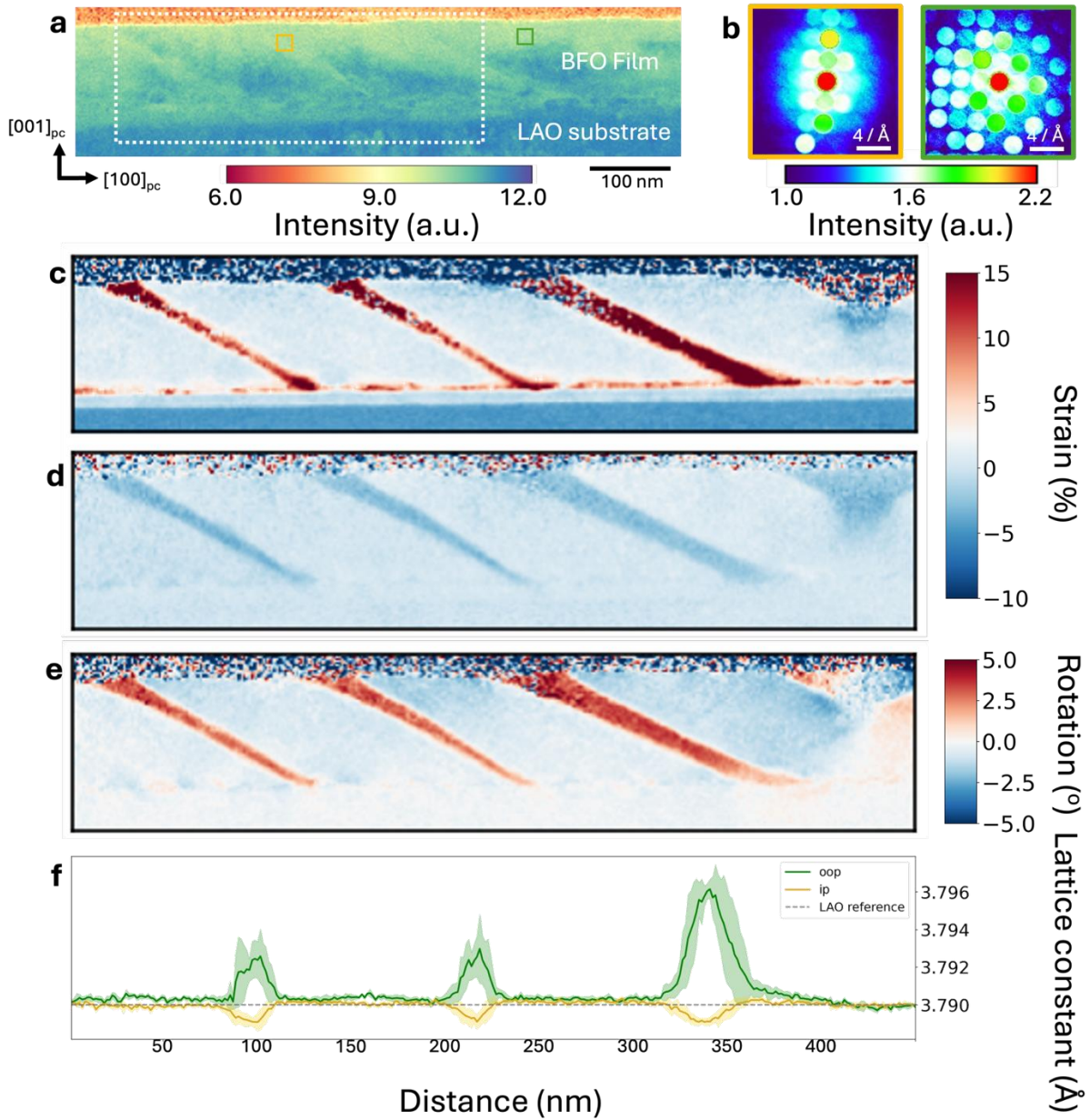

**Figure 3. Lattice strain variation within the MPB probed by cross-sectional scanning electron nanodiffraction.** (a) Virtual annular dark-field STEM image of the MPB region. Contrast in the stripe-like phase boundaries indicates the presence of two separate T' and R' phases. (b) The distinct CBED patterns correspond to the yellow and green box regions in the ADF image, showing clear out-of-plane lattice rotation off (yellow) and along (green) the [010]$_{pc}$ zone axis. (c, d) Out-of-plane and in-plane train maps of the MPB cross-section from the white dashed box in (a). The T'-phase regions exhibit exceptionally large strain of ~15% along out-of-plane and ~5% along the in-plane. (e) Lattice rotation map of the same region, indicating ~3.8° lattice rotation from T' to R' domains. (f) Line profiles of both the *a*- and *c*-axes lattice constants variation.

To rationalize the experimentally observed stabilization of ordered R'/T' ($M_A$/$M_C$) twin domains in strained BFO thin films, we performed phase-field simulations based on a LAO/BFO heterostructure model, incorporating mechanical and electrical boundary conditions consistent with the experimental setting (details in Methods). Three representative twin-domain configurations, pure R' phase twins, pure T' phase twins, and ordered R'/T' twin domains, were simulated and compared. **Figures 4a-c** show the relaxed polarization configurations corresponding to these three states.

In the pure $M_A$ and $M_C$ twin configurations **(Figures 4a and 4b)**, ferroelectric variants arrange in an ordered, periodic manner along the in-plane *y* direction, forming well-defined stripe domains separated by nearly planar and coherent twin boundaries. In contrast, the mixed $M_A$/$M_C$ configuration preserves twin ordering within each phase, while the interphase boundaries exhibit a markedly different morphology characterized by spatial intermixing and strongly non-coplanar interfaces, as highlighted by the white dashed lines in **Figure 4c**. The corresponding in-plane strain distributions $\varepsilon_{xx}$ **(Figure S2)** show that strain remains relatively homogeneous within the pure $M_A$ and $M_C$ twin states, whereas pronounced strain localization develops at the twin and interphase boundaries in the mixed $M_A$/$M_C$ configuration. We note that the strain concentration is predominantly associated with the $M_C$-type domains and their adjacent interphase regions, in good agreement with the experimentally observed strain maps. **Figure 4d** compares the Landau, elastic, and total free-energy densities of the three twin-domain configurations, which our calculations reveal to be the dominant terms in the overall free energy. While the pure $M_A$ configuration minimizes the Landau free energy and the pure $M_C$ configuration favors elastic energy relaxation, the mixed $M_A$/$M_C$ state achieves the lowest total free energy by balancing these competing energetic contributions. This competition between Landau and elastic energies provides the thermodynamic basis for the stabilization of mixed-phase twin domains in strained BFO thin films.

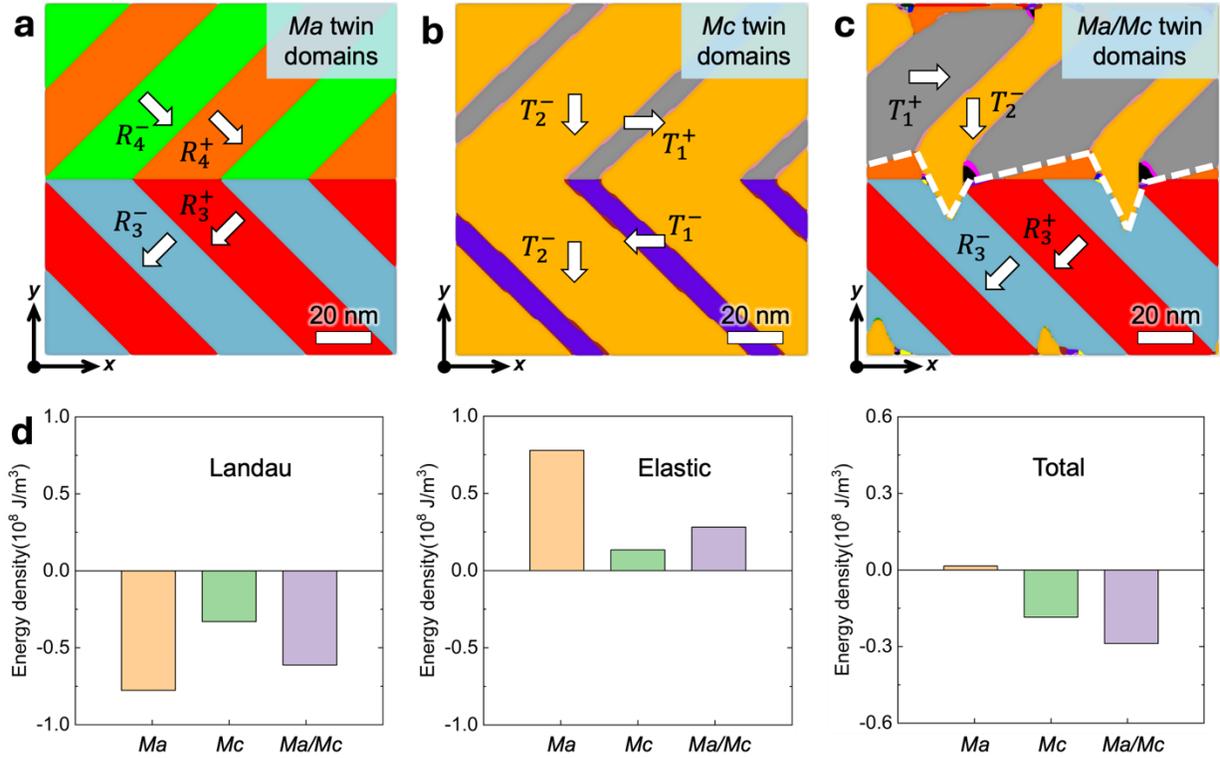

**Figure 4. Phase-field analysis of interphase twin-domain configurations and strain-mediated energetic stabilization in $M_A$, $M_C$, and coexisting $M_A/M_C$ states. (a-c)** Relaxed twin-domain configurations obtained from phase-field simulations, including (a) pure R' ($M_A$) twins, (b) pure T' ($M_C$) twins, and (c) mixed $M_A/M_C$ twin domains. Arrows indicate polarization orientations, and different colors denote distinct ferroelectric variants. **(d)** Comparison of the Landau, elastic, and total free-energy densities for the three domain states.

## 3. Conclusion

In summary, we demonstrate that strain-driven MPBs in BiFeO$_3$ thin films can self-organize into laterally quasi-periodic, phase-separated regions with ~20 m intervals across a 3 mm-wide sample, bounded by millimeter-long MPBs together with a previously unreported class of zigzag interphase boundaries. This mesoscopic ordering of alternating interphase boundaries was resolved using diffraction-based electron microscopy, specifically DREDI and EBSD. Reliable phase discrimination and large-area mapping were enabled by detecting distinct signatures in the Kikuchi intensities of the R' and T' phases.

The correlative atomic-scale MEP structural measurements, nanoscale SEND strain mapping, and mesoscale EBSD analysis establish that these interphase boundaries are associated with laterally modulated strain fields rather than a uniform strain state. Phase-field simulations further show that the mixed R'/T' configuration exhibits the lowest total free energy compared to the pure phase twin states, resulting from the balance between Landau and elastic energy contributions. From a technological perspective, engineering the ordering of nanoscale functional interphase boundaries across macroscopic length scales represents an important step toward scalable strain-engineered ferroic devices.

## 4. Experimental Section/Methods

**Pulsed-Laser Deposition of Thin-Film Heterostructures**
All thin films are prepared by PLD using a 248nm KrF excimer laser (COMPex-Pro, Coherent) in an on-axis geometry with a target-substrate distance of 5cm. We clean $LaAlO_3$ substrates by sonication in Acetone and Isopropyl Alcohol for 5 minutes each followed by DI water treatment and fix the substrates to a heater using thermally conductive silver paint. We achieve a base pressure of $5 \times 10^{-6}$ Torr for both the LSMO and BFO. LSMO was used as a buffer layer and deposited at substrate temperature of 700 °C, fluence of 1.2 $J/cm^2$, $O_2$ pressure of 100 mTorr, and pulse rate of 4 Hz. During the BFO we introduce a dynamic $O_2$ environment of 150 mTorr, and pulse the laser on the target with 10 Hz repetition rate at a fluence of 0.1.3 $J/cm^2$. Following deposition of the BFO, the films were cooled in 300 Torr of oxygen to room temperature.

**Ferroelectric Domain Characteristics**
Piezoelectric force microscopy (PFM) imaging is done using the MFP-3D, Asylum Research AFM. All measurements use a silicon cantilever coated with Pt as the conducting top electrode for local application of an electric field.

**EBSD**
Electron backscattered diffraction (EBSD) datasets were acquired under standard EBSD geometry, with the sample mounted on a 70 ° pre-tilt stage to enhance collection efficiency on a side-entry EBSD detector. The measurements were conducted using Thermo Scientific Helios G4 UXe PFIB/SEM equipped with an Oxford Instruments Symmetry S2 EBSD detector. The datasets were collected using an accelerating voltage of 10 kV at a beam current of 1.6 nA. The real-space collection step size is 10-30 nm with diffraction space resolution of 622 by 532 pixels. Dynamical diffraction simulations of Kikuchi patterns were performed using AztecCrystal software, employing the Bloch-wave algorithm.

**STEM and MEP**
The cross-sectional STEM samples were prepared using the standard FIB lift-out process in a Thermo Fisher Helios G5 Ga FIB-SEM DualBeam system. Sequential carbon and platinum depositions using the electron beam were applied to prevent ion-beam-induced damage of the thin film surface. Initial lamella thinning was performed using a 30 kV $Ga^+$ ion beam until a thickness of approximately 100 nm. Low energy milling was conducted using a 5 kV ion beam to reach a final thickness of 30 nm. A final polishing step using a 2 kV ion beam was applied to remove surface amorphization and minimize residual beam damage.
STEM imaging was performed using an aberration-corrected Thermo Fisher Spectra 200 X-CFEG STEM operated at 200 keV, with a probe semi-convergence angle of 30 mrad. High-angle annular dark field (HAADF) images were acquired with 100 pA probe current and a dwell time of 200 ns per pixel. Rigid registration was applied to 50 sequential frames to correct for the stage drift. MEP datasets were collected using a Dectric ARINA direct electron detector. The dataset had a total of 512 by 512 pixels probe positions, with a collection time of 35 s per pixel. The real-space step size was ~0.372 Å, yielding a total dose of ~1.6 × 10⁶ electrons/Å². MEP reconstruction was done using the PtyRAD package[32] based on the automatic differentiation minimization algorithm. The reconstruction was performed using 25 slices, each 8 Å thick, and ten probe modes. The presented results were obtained after 1000 iterations. Atomic positions were extracted using center-of-mass refinement and 2D Gaussian fitting implemented in the Atomap package[33]. A custom Python script was used to perform further analysis on the polarization, tetragonality, and unit cell rotation of the BFO thin film.

**Phase-field simulations**

In our phase-field framework, two coupled sets of order parameters are considered: the spontaneous ferroelectric polarization $P_i$ (i = 1-3) and spontaneous oxygen octahedral tilt (OT) $\theta_i$ (i = 1-3). The temporal evolution of these order parameters in BFO thin films grown on LSMO-buffered LAO substrate is governed by the following time-dependent Ginzburg-Landau (TDGL) equation[34]:

$$\frac{\partial \phi}{\partial t} = -L \frac{\delta F}{\delta \phi}, \quad (1)$$

where $\phi$ represents the order parameter (either $P_i$ or $\theta_i$), $t$ is the time, and $L$ denotes the kinetic coefficient associated with domain-wall mobility. The total free energy $F$ consists of bulk, gradient, elastic, and electrostatic contributions, and is expressed as a volume integral:

$$F = \int (f_{bulk}(P_i, \theta_i) + f_{grad}(P_{i,j}, \theta_{i,j}) + f_{elas}(P_i, \theta_i, \varepsilon_{ij}) + f_{elec}(P_i, E_i)) \, dV. \quad (2)$$

The bulk energy density takes the form:

$$f_{bulk} = \alpha_{ij} P_i P_j + \alpha_{ijkl} P_i P_j P_k P_l + \alpha_{ijklmn} P_i P_j P_k P_l P_m P_n + \alpha_{ijklmnuv} P_i P_j P_k P_l P_m P_n P_u P_v + \beta_{ij} \theta_i \theta_j + \beta_{ijkl} \theta_i \theta_j \theta_k \theta_l + t_{ijkl} P_i P_j \theta_k \theta_l, \quad (3)$$

where $\alpha_{ij}$, $\alpha_{ijkl}$, $\alpha_{ijklmn}$, $\alpha_{ijklmnuv}$, $\beta_{ij}$, and $\beta_{ijkl}$ are the Landau energy coefficients of polarization and OT order parameters, while $t_{ijkl}$ describes their coupling.

The gradient energy density is given by:

$$f_{grad} = \frac{1}{2} g_{ijkl} P_{i,j} P_{k,l} + \frac{1}{2} h_{ijkl} \theta_{i,j} \theta_{k,l}, \quad (4)$$

where $g_{ijkl}$ and $h_{ijkl}$ are gradient energy coefficients.

The elastic energy density can be written as:

$$f_{elas} = \frac{1}{2} c_{ijkl} (\varepsilon_{ij} - \varepsilon_{ij}^0)(\varepsilon_{kl} - \varepsilon_{kl}^0), \quad (5)$$

where $c_{ijkl}$ denotes the elastic stiffness tensor and $\varepsilon_{ij}$ is the total strain. The eigenstrain $\varepsilon_{ij}^0$ arises from the coupling between the strain, polarization, and OT order parameters, and is expressed as $\varepsilon_{ij}^0 = Q_{ijkl} P_k P_l + \Lambda_{ijkl} \theta_k \theta_l$, with $Q_{ijkl}$ and $\Lambda_{ijkl}$ being the corresponding coupling coefficients.

The electrostatic energy density is defined as:

$$f_{elec} = -\frac{1}{2} \kappa_0 \kappa_b E_i E_j - E_i P_i, \quad (6)$$

where $\kappa_0$ and $\kappa_b$ are the vacuum and relative permittivity, respectively, and $E_i$ represents the total electric field.

A three-dimensional simulation with system size of $128\Delta x \times 128\Delta y \times 50\Delta z$ grid points was performed. The grid spacing is chosen as $\Delta x = \Delta y = \Delta z = 1.0$ nm. The system consists of 10 grids of LAO substrate, 20 grids of BFO thin film and 10 grids of air layer from the bottom to the top along the thickness direction. Periodic boundary conditions were applied along the in-plane directions, whereas a superposition method was adopted along the out-of-plane direction[35]. The mechanical boundary conditions were specified such that the out-of-plane stress is fully relaxed at the film surface, while the out-of-plane displacement is fixed to zero at the bottom of the substrate, sufficiently far from the film/substrate interface. A short-circuit electrical boundary condition was imposed throughout the simulations. The initial polarization configurations were constructed based on experimentally observed domain morphologies. Specifically, three representative types of twin-domain configurations were designed: (i) pure *Ma*-phase *R*-like twin domains, (ii) pure *Mc*-phase *T*-like twin domains, and (iii) mixed *Ma*/*Mc* twin domains arranged in an alternating sequence along the *y* direction. Small-amplitude random perturbations were subsequently added to these structures to mimic thermal fluctuations. All material parameters for BFO are available in the literatures[18,36] and listed in **Table S1**. The simulations were performed at a temperature of 298 K with a normalized time step of 0.01.

## Acknowledgements

The authors acknowledge fruitful discussions with Prof. Zijian Hong and Prof. Sergei Kalinin. This work was primarily supported by the U.S. Department of Energy, Basic Energy Sciences under award No. DE-SC0025423 (T.-R.L., Y.-T.S.). Electron microscopy studies were performed at the Core Center of Excellence in Nano Imaging at the University of Southern California. The authors thank A. Avishai, C. Marks, H. Khant, P. Buenconsejo, and J. Curulli for technical support and careful maintenance of the instruments. S.H., M.R., D.G.S., and R.R. acknowledge support from the Army Research Office under the ETHOS MURI via cooperative agreement W911NF-21-2-0162 and the Army Research Laboratory under Cooperative Agreement Number W911NF-24-2-0100. The views and conclusions contained in this document are those of the authors and should not be interpreted as representing the official policies, either expressed or implied, of the Army Research Laboratory or the U.S. Government. The U.S. Government is authorized to reproduce and distribute reprints for Government purposes, notwithstanding any copyright notation herein. P.G. and R.R. acknowledge Kepler Computing for providing funds for this research work.

## Data Availability Statement

Raw data are available from the corresponding author upon reasonable request.

# References

[1] B. Noheda, J. A. Gonzalo, L. E. Cross, R. Guo, S.-E. Park, D. E. Cox, G. Shirane, *Phys. Rev. B* **2000**, *61*, 8687.
[2] S.-E. Park, T. R. Shrout, *J. Appl. Phys.* **1997**, *82*, 1804.
[3] R. Ghanbari, H. Kp, K. Patel, H. Zhou, T. Zhou, R. Liu, L. Wu, A. Khandelwal, K. J. Crust, S. Hazra, J. Carroll, C. J. G. Meyers, J. Wang, S. Prosandeev, H. Qiao, Y.-H. Kim, Y. Nabei, M. Chi, D. Sun, N. Balke, M. Holt, V. Gopalan, J. E. Spanier, D. A. Muller, L. Bellaiche, H. Y. Hwang, R. Xu, *Nat. Commun.* **2025**, *16*, 7766.
[4] R. J. Zeches, M. D. Rossell, J. X. Zhang, A. J. Hatt, Q. He, C.-H. Yang, A. Kumar, C. H. Wang, A. Melville, C. Adamo, G. Sheng, Y.-H. Chu, J. F. Ihlefeld, R. Erni, C. Ederer, V. Gopalan, L. Q. Chen, D. G. Schlom, N. A. Spaldin, L. W. Martin, R. Ramesh, *Science* **2009**, *326*, 977.
[5] J. Wang, J. B. Neaton, H. Zheng, V. Nagarajan, S. B. Ogale, B. Liu, D. Viehland, V. Vaithyanathan, D. G. Schlom, U. V. Waghmare, N. A. Spaldin, K. M. Rabe, M. Wuttig, R. Ramesh, *Science* **2003**, *299*, 1719.
[6] O. Diéguez, O. E. González-Vázquez, J. C. Wojdeł, J. Íñiguez, *Phys. Rev. B* **2011**, *83*, 094105.
[7] Switching the spin cycloid in BiFeO3 with an electric field | Nature Communications, .
[8] D. Vanderbilt, M. H. Cohen, *Phys. Rev. B* **2001**, *63*, 094108.
[9] S. R. Burns, O. Paull, R. Bulanadi, C. Lau, D. Sando, J. M. Gregg, N. Valanoor, *Phys. Rev. Mater.* **2021**, *5*, 034404.
[10] Q. He, Y.-H. Chu, J. T. Heron, S. Y. Yang, W. I. Liang, C. Y. Kuo, H. J. Lin, P. Yu, C. W. Liang, R. J. Zeches, W. C. Kuo, J. Y. Juang, C. T. Chen, E. Arenholz, A. Scholl, R. Ramesh, *Nat. Commun.* **2011**, *2*, 225.
[11] D. Sando, F. Appert, B. Xu, O. Paull, S. R. Burns, C. Carrétéro, B. Dupé, V. Garcia, Y. Gallais, A. Sacuto, M. Cazayous, B. Dkhil, J. M. Le Breton, A. Barthélémy, M. Bibes, L. Bellaiche, V. Nagarajan, J. Juraszek, *Appl. Phys. Rev.* **2019**, *6*, 041404.
[12] C.-E. Cheng, H.-J. Liu, F. Dinelli, Y.-C. Chen, C.-S. Chang, F. S.-S. Chien, Y.-H. Chu, *Sci. Rep.* **2015**, *5*, 8091.
[13] J. Zhang, X. Ke, G. Gou, J. Seidel, B. Xiang, P. Yu, W.-I. Liang, A. M. Minor, Y. Chu, G. Van Tendeloo, X. Ren, R. Ramesh, *Nat. Commun.* **2013**, *4*, 2768.
[14] Y. Heo, J. Hong Lee, L. Xie, X. Pan, C.-H. Yang, J. Seidel, *NPG Asia Mater.* **2016**, *8*, e297.
[15] L. Caretta, Y.-T. Shao, J. Yu, A. B. Mei, B. F. Grosso, C. Dai, P. Behera, D. Lee, M. McCarter, E. Parsonnet, H. Kp, F. Xue, X. Guo, E. S. Barnard, S. Ganschow, Z. Hong, A. Raja, L. W. Martin, L.-Q. Chen, M. Fiebig, K. Lai, N. A. Spaldin, D. A. Muller, D. G. Schlom, R. Ramesh, *Nat. Mater.* **2023**, *22*, 207.
[16] Y. L. Li, S. Y. Hu, Z. K. Liu, L. Q. Chen, *Appl. Phys. Lett.* **2002**, *81*, 427.
[17] H. M. Christen, J. H. Nam, H. S. Kim, A. J. Hatt, N. A. Spaldin, *Phys. Rev. B* **2011**, *83*, 144107.
[18] F. Xue, Y. Li, Y. Gu, J. Zhang, L.-Q. Chen, *Phys. Rev. B* **2016**, *94*, 220101.
[19] A. J. Hatt, N. A. Spaldin, C. Ederer, *Phys. Rev. B* **2010**, *81*, 054109.
[20] H. Kp, R. Xu, K. Patel, K. J. Crust, A. Khandelwal, C. Zhang, S. Prosandeev, H. Zhou, Y.-T. Shao, L. Bellaiche, H. Y. Hwang, D. A. Muller, *Nat. Mater.* **2025**, *24*, 1433.
[21] Z. Chen, Y. Jiang, Y.-T. Shao, M. E. Holtz, M. Odstrčil, M. Guizar-Sicairos, I. Hanke, S. Ganschow, D. G. Schlom, D. A. Muller, *Science* **2021**, *372*, 826.
[22] A. R. Damodaran, C.-W. Liang, Q. He, C.-Y. Peng, L. Chang, Y.-H. Chu, L. W. Martin, *Adv. Mater.* **2011**, *23*, 3170.

[23] T.-R. Liu, K. Jagadish, X. Guo, M. Ramesh, P. Meisenheimer, H. Kumarasubramanian, S. Husain, A. V. Ngo, A. Avishai, J. Ravichandran, D. G. Schlom, R. Ramesh, Y.-T. Shao, *Revealing buried ferroelectric topologies by depth-resolved electron diffraction imaging.*
[24] P. G. Callahan, M. De Graef, *Microsc. Microanal.* **2013**, *19*, 1255.
[25] J. M. Zuo, J. C. H. Spence, *Advanced Transmission Electron Microscopy*, Springer, New York, NY, **2017**.
[26] B. Fultz, J. Howe, *Transmission Electron Microscopy and Diffractometry of Materials*, Springer, Berlin, Heidelberg, **2013**.
[27] A. Winkelmann, C. Trager-Cowan, F. Sweeney, A. P. Day, P. Parbrook, *Ultramicroscopy* **2007**, *107*, 414.
[28] E. Padgett, M. E. Holtz, P. Cueva, Y.-T. Shao, E. Langenberg, D. G. Schlom, D. A. Muller, *Ultramicroscopy* **2020**, *214*, 112994.
[29] Y. L. Tang, Y. L. Zhu, M. J. Zou, Y. J. Wang, X. L. Ma, *J. Appl. Phys.* **2021**, *129*, 184101.
[30] K. M. Holsgrove, M. Duchamp, M. S. Moreno, N. Bernier, A. B. Naden, J. G. M. Guy, N. Browne, A. Gupta, J. M. Gregg, A. Kumar, M. Arredondo, *RSC Adv.* **2020**, *10*, 27954.
[31] J. Sapriel, *Phys. Rev. B* **1975**, *12*, 5128.
[32] C.-H. Lee, S. E. Zeltmann, D. Yoon, D. Ma, D. A. Muller, *Microsc. Microanal.* **2025**, *31*, ozaf070.
[33] M. Nord, P. E. Vullum, I. MacLaren, T. Tybell, R. Holmestad, *Adv. Struct. Chem. Imaging* **2017**, *3*, 9.
[34] L.-Q. Chen, *J. Am. Ceram. Soc.* **2008**, *91*, 1835.
[35] L. Q. Chen, J. Shen, *Comput. Phys. Commun.* **1998**, *108*, 147.
[36] F. Xue, Y. Gu, L. Liang, Y. Wang, L.-Q. Chen, *Phys. Rev. B* **2014**, *90*, 220101.

## Supporting Information

Supporting Information is available from the Wiley Online Library or from the author.